\documentclass[12pt]{article}

\usepackage{graphicx}
\usepackage{amsmath}
\usepackage{amssymb}
\allowdisplaybreaks
\numberwithin{equation}{section}

\newcommand{\cS}{{\mathcal S}}
\newcommand{\cR}{{\mathcal R}}
\newcommand{\cC}{{\mathcal C}}
\newcommand{\cJ}{{\mathcal J}}
\newcommand{\fL}{{\mathfrak L}}
\newcommand{\fR}{{\mathfrak R}}
\newcommand{\fQ}{{\mathfrak Q}}
\newcommand{\fS}{{\mathfrak S}}
\newcommand{\fC}{{\mathfrak C}}
\newcommand{\fI}{{\mathfrak I}}
\newcommand{\fU}{{\mathfrak U}}
\newcommand{\fP}{{\mathfrak P}}
\newcommand{\fK}{{\mathfrak K}}
\newcommand{\alg}[1]{\mathfrak{#1}}

\begin{document}

\baselineskip=16pt plus 0.2pt minus 0.1pt

\begin{titlepage}
\title{
\hfill\parbox{4cm}
{\normalsize MIT-CTP-3853}\\
\vspace{1cm}
{\Large\bf A Secret Symmetry of the AdS/CFT S-matrix}
}
\author{
{\sc Takuya Matsumoto}$^{1}$\,{}\thanks
{{\tt m05044c@math.nagoya-u.ac.jp}}\\[6pt]
{\sc Sanefumi Moriyama}${}^{1,2}$\,{}\thanks
{{\tt moriyama@math.nagoya-u.ac.jp}}\\[6pt]
{\sc Alessandro Torrielli}${}^{2}$\,{}\thanks
{{\tt torriell@mit.edu}}\\[12pt]
${}^{1}${\it Graduate School of Mathematics, Nagoya University,}\\
{\it Nagoya 464-8602, Japan}\\[6pt]
${}^{2}${\it Center for Theoretical Physics, 
Massachusetts Institute of Technology,}\\
{\it Cambridge, MA02139, USA}\\
}
\date{\normalsize August, 2007}
\maketitle
\thispagestyle{empty}

\begin{abstract}
\normalsize
We find a new quantum Yangian symmetry of the AdS/CFT S-matrix, which
complements the original $\alg{su(2|2)}$ symmetry to $\alg{gl(2|2)}$ and
does not have a Lie algebra analog.
Our finding is motivated by the Yangian double structure discovered at
the classical level.
\end{abstract}

\end{titlepage}

\section{Introduction}

One of the most important results in the study of the integrable spin
chain inspired by AdS/CFT \cite{MZ}\footnote{We refer the reader to the
reviews \cite{revs} and references therein.} is that the relevant
S-matrix \cite{BeiSmat} can be determined uniquely by the
$\alg{su(2|2)}$ Lie algebra symmetry of the problem, up to an overall
dressing factor.
This fact reduces the problem of the dynamics into a single function.
The dressing factor satisfies the crossing symmetry constraint
\cite{Jan} originating from the underlying Hopf algebra structure
\cite{Jan,hopf,PST}.
A remarkable solution to Janik's equation, reproducing the asymptotic
behavior in the weak and strong coupling region \cite{pert,AF}, was
recently proposed by \cite{BES} and passed highly non-trivial checks
\cite{BBKS}.

In order to gain a deeper understanding of the hidden algebra
responsible for such a structure\footnote{See \cite{Thomas} for a recent
test of integrability in the near-flat space limit \cite{MS}.}, it is
desirable to understand all the symmetries of the model
\cite{symm,FZ,BZ}.
In \cite{BeisYang} it was shown that the whole Lie algebra symmetry
$\alg{su(2|2)}$ is lifted to the infinite-dimensional Yangian symmetry
by generalizing the standard formula of Drinfeld's first realization of
Yangians \cite{Dr1}\footnote{Note that we can freely rescale $\hbar$ at
the present stage. We shall fix it later to a convenient value.}:
\begin{align}
\Delta\widehat\cJ^A=\widehat\cJ^A\otimes 1
+1\otimes\widehat\cJ^A
+\frac{i}{2}\hbar f^A_{BC}\cJ^B\otimes\cJ^C~,\quad
{\rm S}(\widehat\cJ^A)=-\widehat\cJ^A
+\frac{i}{4}\hbar f^A_{BC}f^{BC}_D\cJ^D~.
\label{DandS}
\end{align}
Since the Cartan matrix which is used to raise and lower the indices is
degenerate, one appealed to $\alg{sl(2)}$ automorphisms \cite{Beinlin}
which couple to the central charges.

In the study of the classical (near BMN) limit \cite{Tor} it was noticed
that one of the poles reveals the Casimir operator of the Lie algebra
$\alg{gl(2|2)}$, instead of the expected symmetry algebra
$\alg{su(2|2)}$.
Although this fact suggests that the model has a larger symmetry of
$\alg{gl(2|2)}$, clearly the additional generator $\fI$ which extends
$\alg{su(2|2)}$ to $\alg{gl(2|2)}$
\begin{align}
\fI|\phi^a\rangle=I|\phi^a\rangle~,\quad
\fI|\psi^\alpha\rangle=-I|\psi^\alpha\rangle~,
\label{fI}
\end{align}
is not a symmetry of the S-matrix, when equipped with a trivial
coproduct.

A possible resolution to this problem was subsequently found in
\cite{MorTor} by rewriting the classical r-matrix in the form of a
Yangian double (Drinfeld's second realization \cite{Dr2}).
There, one found it necessary to include an infinite family of
generators $\fI_n$ to be able to factorize the classical r-matrix.
Although the coefficient of the additional generators $\fI_n$ remains
non-trivial for the infinite-dimensional Yangian algebra, it actually
vanishes for the classical Lie algebra at level $n=0$.

Here we would like to continue the study of this additional symmetry.
Since it does not have a Lie algebraic analog, we will call it a secret
symmetry.
Starting from the first higher Yangian level, the coproducts are usually
non-trivial, and there is a chance for the new generators to be
symmetries.
In fact, we will find a quantum Yangian generator proportional to $\fI$,
equipped with a coproduct that makes it a symmetry of the S-matrix.
We shall exploit a hybrid version of the arguments adopted in
\cite{BeisYang} and \cite{MorTor}.
Namely, we apply the formulas \eqref{DandS} together with the fact that
the $\fI$ operator couples to the central charge $\fC$ \cite{MorTor}.
We find
\begin{align}
\Delta\widehat\fI=\widehat\fI\otimes 1+1\otimes\widehat\fI
+\frac{i}{2g}\bigl(\fQ^\alpha{}_a\fU^{-1}\otimes\fS^a{}_\alpha
+\fS^a{}_\alpha\fU^{+1}\otimes\fQ^\alpha{}_a\bigr)~,\quad
{\rm S}(\widehat\fI)=-\widehat\fI+\frac{2i}{g}\fC~,\label{SI}
\end{align}
where $\widehat\fI$ acts as $\fI$ in \eqref{fI}, with the eigenvalue $I$
replaced by $\widehat I$. In the main text we will show that the
coproduct $\Delta\widehat\fI$ is a symmetry of the S-matrix if we choose
\begin{align}
\widehat I=\frac{1}{4}(x^++x^--1/x^+-1/x^-)~.
\end{align}
The Hopf algebra structure associated to the
Lie algebra generators $\fQ^\alpha{}_a$, $\fS^a{}_\alpha$ and $\fC$ in
\eqref{SI} is as described in \cite{BeisYang}.
The braiding factor $\fU$ is a central element with eigenvalue
$U=\sqrt{x^+/x^-}$.
As from \eqref{DandS}, it is easy to check that the defining relation of
Hopf algebras $\mu\circ({\rm S}\otimes 1)\circ\Delta=\eta\circ\epsilon$
($\mu$ is the algebra multiplication) holds for $\widehat\fI$, with the
counit $\epsilon(\widehat\fI)=0$.

A natural question arises whether the generator $\widehat\fI$ we found
is a generator in Drinfeld's second realization, so that we can use it
directly to construct the universal R-matrix.
Taking the classical limit of this operator, which essentially consists
in replacing $x^\pm$ by the classical spectral parameter $x$
\cite{AF,Tor}, we find
\begin{align}
\widehat I\to\frac{1}{2}(x-1/x)~.
\end{align}
This coincides with the expression of the first level generator
$\fI_{n=1}$ found in \cite{MorTor}, where Drinfeld's second realization
was employed.
Therefore, it appears natural to expect that the operator $\widehat\fI$
is in the second realization.
However, as we will argue below, this is not likely to be the case.

Since the generators in Drinfeld's second realization are the ones
directly appearing in the universal form of the R-matrix, they have to
be consistent with the crossing equation found by Janik \cite{Jan}:
\begin{align}
({\cal{C}}^{-1}\otimes 1)\bigl[\cR(1/x_1^\pm,x_2^\pm)\bigr]^{st_1}
({\cal{C}}\otimes 1)\cR(x_1^\pm,x_2^\pm)=1~,
\label{Janiks}
\end{align}
where ${\cal{C}}$ is a (bosonic) charge conjugation matrix and $st_1$
denotes supertransposition in the first entry.
This equation originates from combining the knowledge of the crossing
symmetry transformation of the Lie algebra ($n=0$) generators with the
fundamental property $({\rm S}\otimes 1){\cal{R}}={\cal{R}}^{-1}$ of
quasi-triangular Hopf algebras.
On the other hand, if we assume that the universal R-matrix $\cal{R}$
admits some expansion in powers of the Yangian generators $\cJ_n$,
\eqref{Janiks} leads to the expectation that all $\cJ_n$'s have to
satisfy the antipode relation
\begin{align}
\label{antipoderel}
{\rm S}(\cJ_n(x^\pm))
={\cC}^{-1}\bigl[\cJ_n(1/x^\pm)\bigr]^{st}{\cC}~,
\end{align}
with one and the same charge conjugation $\cal{C}$.
Apparently, although the operator $\widehat\fI$ satisfies the relation
$\cC^{-1}[\widehat\fI(1/x^\pm)]^{st}\cC=-\widehat\fI(x^\pm)$, it does
not satisfy \eqref{antipoderel}.
Hence, it is difficult to expect this operator to be in the second 
realization\footnote{Nevertheless, we still find it a little bizarre
that \eqref{antipoderel} is not satisfied by this new symmetry. This
issue certainly deserves further investigation, on which we reserve to
come back in the future.}.

As a side remark, we notice that all the generators found in
\cite{MorTor} satisfy the classical antipode relation
\begin{align}
{\rm S}(\cJ_n(x))
={\cC}_0^{-1}\bigl[\cJ_n(1/x)\bigr]^{st}{\cC}_0~,
\label{clantipode}
\end{align}
where $\cC_0$ is the classical charge conjugation, and the classical
antipode is given by
\begin{equation}
{\rm S}(\cJ_n(x))=-\cJ_n(x)~. 
\end{equation} 
The explicit expression of $\cC_0$ is not necessary to show that the
classical antipode relation \eqref{clantipode} holds for the infinite
tower of generators, provided the relation is satisfied by the Lie
algebra generators at level $n=0$ \cite{Jan,PST,FZ}.

In the last part of the introduction, let us recall for convenience the
action of the supercharges\footnote{We follow the notation of
\cite{Beinlin}.
In particular, note that $g$ in \cite{Beinlin} is related to the one in
\cite{BeiSmat} by $g_{\cite{Beinlin}}=g_{\cite{BeiSmat}}/\sqrt{2}$.}
\begin{align}
\fQ^\alpha{}_a|\phi^b\rangle=a\delta^b_a|\psi^\alpha\rangle~,&\quad
\fQ^\alpha{}_a|\psi^\beta\rangle
=b\epsilon^{\alpha\beta}\epsilon_{ab}|\phi^b\rangle~,\nonumber\\
\fS^a{}_\alpha|\phi^b\rangle
=c\epsilon^{ab}\epsilon_{\alpha\beta}|\psi^\beta\rangle~,&\quad
\fS^a{}_\alpha|\psi^\beta\rangle=d\delta^\beta_\alpha|\phi^a\rangle~,
\end{align}
with $a$, $b$, $c$ and $d$ parameterized by
\begin{align}
a=\sqrt{g}\gamma~,\quad
b=\sqrt{g}\frac{\alpha}{\gamma}\biggl(1-\frac{x^+}{x^-}\biggr)~,\quad
c=\sqrt{g}\frac{i\gamma}{\alpha x^+}~,\quad
d=\sqrt{g}\frac{x^+}{i\gamma}\biggl(1-\frac{x^-}{x^+}\biggr)~,
\end{align}
and the R-matrix \cite{BeiSmat,Beinlin} that we will use in this paper:
\begin{align}
\cR_{12}|\phi^a_1\phi^b_2\rangle
&=\frac{1}{2}(A_{12}-B_{12})|\phi^a_1\phi^b_2\rangle
+\frac{1}{2}(A_{12}+B_{12})|\phi^b_1\phi^a_2\rangle
+\frac{1}{2}C_{12}\epsilon^{ab}\epsilon_{\alpha\beta}
|\psi^\alpha_1\psi^\beta_2\rangle~,\nonumber\\
\cR_{12}|\psi^\alpha_1\psi^\beta_2\rangle
&=-\frac{1}{2}(D_{12}-E_{12})|\psi^\alpha_1\psi^\beta_2\rangle
-\frac{1}{2}(D_{12}+E_{12})|\psi^\beta_1\psi^\alpha_2\rangle
-\frac{1}{2}F_{12}\epsilon^{\alpha\beta}\epsilon_{ab}
|\phi^a_1\phi^b_2\rangle~,\nonumber\\
\cR_{12}|\phi^a_1\psi^\beta_2\rangle
&=G_{12}|\phi^a_1\psi^\beta_2\rangle
+H_{12}|\psi^\beta_1\phi^a_2\rangle~,\nonumber\\
\cR_{12}|\psi^\alpha_1\phi^b_2\rangle
&=L_{12}|\psi^\alpha_1\phi^b_2\rangle
+K_{12}|\phi^b_1\psi^\alpha_2\rangle~.
\end{align}
The functions $A_{12},B_{12},\ldots$ are given by
\begin{align}
\label{func}
&A_{12}=\frac{x_2^+-x_1^-}{x_2^--x_1^+}~,\quad
B_{12}=\frac{x_2^+-x_1^-}{x_2^--x_1^+}
\biggl(1-2\frac{1-1/x_1^+x_2^-}{1-1/x_1^+x_2^+}
\frac{x_2^--x_1^-}{x_2^+-x_1^-}\biggr)~,\nonumber\\
&\qquad C_{12}=\frac{2\gamma_1\gamma_2U_2}{\alpha x_1^+x_2^+}
\frac{1}{1-1/x_1^+x_2^+}
\frac{x_2^--x_1^-}{x_2^--x_1^+}~,\nonumber\\
&D_{12}=-\frac{U_2}{U_1}~,\quad
E_{12}=-\frac{U_2}{U_1}
\biggl(1-2\frac{1-1/x_1^-x_2^+}{1-1/x_1^-x_2^-}
\frac{x_2^+-x_1^+}{x_2^--x_1^+}\biggr)~,\nonumber\\
&\qquad F_{12}=-\frac{2\alpha(x_1^+-x_1^-)(x_2^+-x_2^-)}
{\gamma_1\gamma_2U_1x_1^-x_2^-}
\frac{1}{1-1/x_1^-x_2^-}
\frac{x_2^+-x_1^+}{x_2^--x_1^+}~,\nonumber\\
&G_{12}=\frac{1}{U_1}\frac{x_2^+-x_1^+}{x_2^--x_1^+}~,\quad
H_{12}=\frac{\gamma_1U_2}{\gamma_2U_1}
\frac{x_2^+-x_2^-}{x_2^--x_1^+}~,\nonumber\\
&L_{12}=U_2\frac{x_2^--x_1^-}{x_2^--x_1^+}~,\quad
K_{12}=\frac{\gamma_2}{\gamma_1}
\frac{x_1^+-x_1^-}{x_2^--x_1^+}~,
\end{align}
where $U=\sqrt{x^+/x^-}$ and $x^\pm$ satisfy the relation
$x^++1/x^+-x^--1/x^-=i/g$.

\section{Secret symmetry}

In this section, we would like to show that the coproduct of the
additional operator
\begin{align}
\Delta\widehat\fI=\widehat\fI\otimes 1+1\otimes\widehat\fI
+\frac{i}{2g}\bigl(\fQ^\alpha{}_a\fU^{-1}\otimes\fS^a{}_\alpha
+\fS^a{}_\alpha\fU^{+1}\otimes\fQ^\alpha{}_a\bigr)~,
\end{align}
is an exact symmetry of the S-matrix:
\begin{align}
\bigl[\Delta\widehat\fI,\cS\bigr]=0~.
\label{comm}
\end{align}
This equation can be expressed in terms of the R-matrix $\cR=\Pi\cS$
($\Pi$ is the graded permutation operator) as
\begin{align}
\bigl[\widehat\fI\otimes 1+1\otimes\widehat\fI,\cR\bigr]
=\frac{i}{2g}\bigl[\{\fQ,\fS\}_\otimes\,\cR
+\cR\,\{\fQ,\fS\}_\otimes\bigr]~.
\label{symmetry}
\end{align}
Here the expression on the right-hand-side is a bookkeeping notation:
\begin{align}
&\{\fQ,\fS\}_\otimes\,\cR
=\bigl(\fQ^\alpha{}_a\otimes\fU^{+1}\fS^a{}_\alpha
+\fS^a{}_\alpha\otimes\fU^{-1}\fQ^\alpha{}_a\bigr)\cR~,\nonumber\\
&\cR\,\{\fQ,\fS\}_\otimes
=\cR\bigl(\fQ^\alpha{}_a\fU^{-1}\otimes\fS^a{}_\alpha
+\fS^a{}_\alpha\fU^{+1}\otimes\fQ^\alpha{}_a\bigr)~.
\end{align}
Note that the braiding factor of the first (second) line is in the
second (first) entry.

The first term on the right-hand-side of \eqref{symmetry} is given by
\begin{align}
&\{\fQ,\fS\}_\otimes\,\cR\,|\phi^a_1\phi^b_2\rangle
=-B_{12}(a_1c_2U_2+c_1a_2U_2^{-1})
\epsilon^{ab}\epsilon_{\alpha\beta}|\psi^\alpha_1\psi^\beta_2\rangle
\nonumber\\&\hspace{5cm}
+C_{12}(-b_1d_2U_2-d_1b_2U_2^{-1})
\bigl(|\phi^a_1\phi^b_2\rangle
-|\phi^b_1\phi^a_2\rangle\bigr)~,\nonumber\\
&\{\fQ,\fS\}_\otimes\,\cR\,|\psi^\alpha_1\psi^\beta_2\rangle
=E_{12}(-b_1d_2U_2-d_1b_2U_2^{-1})
\epsilon^{\alpha\beta}\epsilon_{ab}|\phi^a_1\phi^b_2\rangle
\nonumber\\&\hspace{5cm}
-F_{12}(a_1c_2U_2+c_1a_2U_2^{-1})
\bigl(|\psi^\alpha_1\psi^\beta_2\rangle
-|\psi^\beta_1\psi^\alpha_2\rangle\bigr)~,\nonumber\\
&\{\fQ,\fS\}_\otimes\,\cR\,|\phi^a_1\psi^\beta_2\rangle
=G_{12}(a_1d_2U_2+c_1b_2U_2^{-1})|\psi^\beta_1\phi^a_2\rangle
+H_{12}(-b_1c_2U_2-d_1a_2U_2^{-1})|\phi^a_1\psi^\beta_2\rangle~,
\nonumber\\
&\{\fQ,\fS\}_\otimes\,\cR\,|\psi^\alpha_1\phi^b_2\rangle
=L_{12}(-b_1c_2U_2-d_1a_2U_2^{-1})|\phi^b_1\psi^\alpha_2\rangle
+K_{12}(a_1d_2U_2+c_1b_2U_2^{-1})|\psi^\alpha_1\phi^b_2\rangle~,
\end{align}
while the second term is
\begin{align}
&\cR\,\{\fQ,\fS\}_\otimes\,|\phi^a_1\phi^b_2\rangle
=(a_1U_1^{-1}c_2+c_1U_1a_2)\bigl(E_{12}
\epsilon^{ab}\epsilon_{\alpha\beta}|\psi^\alpha_1\psi^\beta_2\rangle
-F_{12}\bigl(|\phi^a_1\phi^b_2\rangle
-|\phi^b_1\phi^a_2\rangle\bigr)\bigr)~,\nonumber\\
&\cR\,\{\fQ,\fS\}_\otimes\,|\psi^\alpha_1\psi^\beta_2\rangle
=(-b_1U_1^{-1}d_2-d_1U_1b_2)\bigl(-B_{12}
\epsilon^{\alpha\beta}\epsilon_{ab}|\phi^a_1\phi^b_2\rangle
+C_{12}\bigl(|\psi^\alpha_1\psi^\beta_2\rangle
-|\psi^\beta_1\psi^\alpha_2\rangle\bigr)\bigr)~,\nonumber\\
&\cR\,\{\fQ,\fS\}_\otimes\,|\phi^a_1\psi^\beta_2\rangle
=(a_1U_1^{-1}d_2+c_1U_1b_2)
\bigl(L_{12}|\psi^\beta_1\phi^\alpha_2\rangle
+K_{12}|\phi^a_1\psi^\beta_2\rangle\bigr)~,\nonumber\\
&\cR\,\{\fQ,\fS\}_\otimes\,|\psi^\alpha_1\phi^b_2\rangle
=(-b_1U_1^{-1}c_2-d_1U_1a_2)
\bigl(G_{12}|\phi^b_1\psi^\alpha_2\rangle
+H_{12}|\psi^\alpha_1\phi^b_2\rangle\bigr)~.
\end{align}
Using the relations
\begin{align}
&-B_{12}(a_1c_2U_2+c_1a_2U_2^{-1})+E_{12}(a_1U_1^{-1}c_2+c_1U_1a_2)
=2igC_{12}(\widehat I_1+\widehat I_2)~,\nonumber\\
&C_{12}(-b_1d_2U_2-d_1b_2U_2^{-1})-F_{12}(a_1U_1^{-1}c_2+c_1U_1a_2)
=0~,\nonumber\\
&E_{12}(-b_1d_2U_2-d_1b_2U_2^{-1})-B_{12}(-b_1U_1^{-1}d_2-d_1U_1b_2)
=2igF_{12}(\widehat I_1+\widehat I_2)~,\nonumber\\
&-F_{12}(a_1c_2U_2+c_1a_2U_2^{-1})+C_{12}(-b_1U_1^{-1}d_2-d_1U_1b_2)
=0~,\nonumber\\
&G_{12}(a_1d_2U_2+c_1b_2U_2^{-1})+L_{12}(a_1U_1^{-1}d_2+c_1U_1b_2)
=4igH_{12}(\widehat I_1-\widehat I_2)~,\nonumber\\
&H_{12}(-b_1c_2U_2-d_1a_2U_2^{-1})+K_{12}(a_1U_1^{-1}d_2+c_1U_1b_2)
=0~,\nonumber\\
&L_{12}(-b_1c_2U_2-d_1a_2U_2^{-1})+G_{12}(-b_1U_1^{-1}c_2-d_1U_1a_2)
=-4igK_{12}(\widehat I_1-\widehat I_2)~,\nonumber\\
&K_{12}(a_1d_2U_2+c_1b_2U_2^{-1})+H_{12}(-b_1U_1^{-1}c_2-d_1U_1a_2)
=0~,
\label{computation}
\end{align}
if we define $\widehat I_{1(2)}$ as
\begin{align}
\widehat I=\frac{1}{4}(x^++x^--1/x^+-1/x^-)~,
\end{align}
we find
\begin{align}
&\bigl[\{\fQ,\fS\}_\otimes\,\cR+\cR\,\{\fQ,\fS\}_\otimes\bigr]
|\phi^a_1\phi^b_2\rangle
=2igC_{12}(\widehat I_1+\widehat I_2)
\epsilon^{ab}\epsilon_{\alpha\beta}
|\psi^\alpha_1\psi^\beta_2\rangle~,\nonumber\\
&\bigl[\{\fQ,\fS\}_\otimes\,\cR+\cR\,\{\fQ,\fS\}_\otimes\bigr]
|\psi^\alpha_1\psi^\beta_2\rangle
=2igF_{12}(\widehat I_1+\widehat I_2)
\epsilon^{\alpha\beta}\epsilon_{ab}
|\phi^a_1\phi^b_2\rangle~,\nonumber\\
&\bigl[\{\fQ,\fS\}_\otimes\,\cR+\cR\,\{\fQ,\fS\}_\otimes\bigr]
|\phi^a_1\psi^\beta_2\rangle
=4igH_{12}(\widehat I_1-\widehat I_2)
|\psi^\beta_1\phi^a_2\rangle~,\nonumber\\
&\bigl[\{\fQ,\fS\}_\otimes\,\cR+\cR\,\{\fQ,\fS\}_\otimes\bigr]
|\psi^\alpha_1\phi^b_2\rangle
=-4igK_{12}(\widehat I_1-\widehat I_2)
|\phi^b_1\psi^\alpha_2\rangle~.
\label{result}
\end{align}

Most of the computations in \eqref{computation} are tedious but
straightforward, except the first and the third line with $B_{12}$ and
$E_{12}$, which require a more elaborated treatment.
The easiest way to compute them is to separately deal with the `$1$'
term and the `$-2$' term in the parenthesis of $B_{12}$ and $E_{12}$ in
\eqref{func}.
It is not difficult to find that the `$-2$' term is proportional to
$C_{12}$ and $F_{12}$ respectively.
For the `$1$' term, the following identity is useful
\begin{align}
x_2^+x_2^--x_1^+x_1^-=2\frac{x_2^--x_1^-}{1-1/x_1^+x_2^+}
(\widehat I_1+\widehat I_2)~,
\end{align}
which can be obtained by multiplying the identity
\begin{align}
x_2^+x_2^--x_1^+x_1^-=\frac{1}{2}
\bigl((x_2^++x_1^+)(x_2^--x_1^-)+(x_2^-+x_1^-)(x_2^+-x_1^+)\bigr)~,
\end{align}
by the quantity
$(1-1/x_1^+x_2^+)/(x_2^--x_1^-)=(1-1/x_1^-x_2^-)/(x_2^+-x_1^+)$.

Comparing \eqref{result} with
\begin{align}
&\bigl[\widehat\fI\otimes 1+1\otimes\widehat\fI,\cR\bigr]
|\phi^a_1\phi^b_2\rangle
=2(-\widehat I_1-\widehat I_2)(C_{12}/2)
\epsilon^{ab}\epsilon_{\alpha\beta}
|\psi^\alpha_1\psi^\beta_2\rangle~,\nonumber\\
&\bigl[\widehat\fI\otimes 1+1\otimes\widehat\fI,\cR\bigr]
|\psi^\alpha_1\psi^\beta_2\rangle
=2(\widehat I_1+\widehat I_2)(-F_{12}/2)
\epsilon^{\alpha\beta}\epsilon_{ab}
|\phi^a_1\phi^b_2\rangle~,\nonumber\\
&\bigl[\widehat\fI\otimes 1+1\otimes\widehat\fI,\cR\bigr]
|\phi^a_1\psi^\beta_2\rangle
=2(-\widehat I_1+\widehat I_2)H_{12}
|\psi^\beta_1\phi^a_2\rangle~,\nonumber\\
&\bigl[\widehat\fI\otimes 1+1\otimes\widehat\fI,\cR\bigr]
|\psi^\alpha_1\phi^b_2\rangle
=2(\widehat I_1-\widehat I_2)K_{12}
|\phi^b_1\psi^\alpha_2\rangle~,
\end{align}
we eventually find that \eqref{symmetry} holds.

\section{Conclusion}

We have found a new quantum Yangian symmetry of the AdS/CFT S-matrix,
which complements the original $\alg{su(2|2)}$ symmetry to
$\alg{gl(2|2)}$ and does not have a Lie algebra analog.
Using this novel symmetry we can generate several new ones.
For example, we can compute the commutators
$[\Delta\widehat\fI,\Delta\fQ^\alpha{}_{b}]$ and
$[\Delta\widehat\fI,\Delta\fS^a{}_{\beta}]$.
Taking the linear combination with the known invariances of the S-matrix
$\Delta\widehat\fQ^\alpha{}_b$ and $\Delta\widehat\fS^a{}_\beta$
discovered in \cite{BeisYang}, we find new symmetries\footnote{As a
check, we have explicitly verified that one of them, namely
$\Delta\fQ^\alpha{}_{b,1}$, is an exact symmetry of the S-matrix:
$[\Delta\fQ^\alpha{}_{b,1},\cS]=0$.}:
($\hbar=1/(2g)$)
\begin{align}
&\Delta\fQ^\alpha{}_{b,1}
=\fQ^\alpha{}_{b,1}\otimes 1
+\fU^{+1}\otimes\fQ^\alpha{}_{b,1}
+i\hbar
\bigl(\fL^\alpha{}_\gamma\fU^{+1}\otimes\fQ^\gamma{}_b
+\fR^c{}_b\fU^{+1}\otimes\fQ^\alpha{}_c
+\fC\fU^{+1}\otimes\fQ^\alpha{}_b\nonumber\\
&\hspace{7cm}
-\fQ^\gamma{}_b\otimes\fL^\alpha{}_\gamma
-\fQ^\alpha{}_c\otimes\fR^c{}_b
-\fQ^\alpha{}_b\otimes\fC\bigr)~,
\nonumber\\
&\Delta\fQ^\alpha{}_{b,-1}
=\fQ^\alpha{}_{b,-1}\otimes 1
+\fU^{+1}\otimes\fQ^\alpha{}_{b,-1}
+i\hbar\bigl(
-\epsilon^{\alpha\gamma}\epsilon_{bd}\fP\fU^{-1}\otimes\fS^d{}_\gamma
+\epsilon^{\alpha\gamma}\epsilon_{bd}\fS^d{}_\gamma\fU^{+2}\otimes\fP
\bigr)~,
\nonumber\\
&\Delta\fS^a{}_{\beta,1}
=\fS^a{}_{\beta,1}\otimes 1
+\fU^{-1}\otimes\fS^a{}_{\beta,1}
+i\hbar
\bigl(-\fR^a{}_c\fU^{-1}\otimes\fS^c{}_\beta
-\fL^\gamma{}_\beta\fU^{-1}\otimes\fS^a{}_\gamma
-\fC\fU^{-1}\otimes\fS^a{}_\beta\nonumber\\
&\hspace{7cm}+\fS^c{}_\beta\otimes\fR^a{}_c
+\fS^a{}_\gamma\otimes\fL^\gamma{}_\beta
+\fS^a{}_\beta\otimes\fC\bigr)~,
\nonumber\\
&\Delta\fS^a{}_{\beta,-1}
=\fS^a{}_{\beta,-1}\otimes 1
+\fU^{-1}\otimes\fS^a{}_{\beta,-1}
+i\hbar\bigl(
\epsilon^{ac}\epsilon_{\beta\delta}\fK\fU^{+1}\otimes\fQ^\delta{}_c
-\epsilon^{ac}\epsilon_{\beta\delta}\fQ^\delta{}_c\fU^{-2}\otimes\fK
\bigr)~,
\end{align}
where $\fQ^\alpha{}_{b,\pm 1}$ and $\fS^a{}_{\beta,\pm 1}$ are defined
as
\begin{align}
\fQ^\alpha{}_{b,1}=\fQ^\alpha{}_b
\bigl(u\Pi_{\rm b}+v\Pi_{\rm f}\bigr)~,&\quad
\fQ^\alpha{}_{b,-1}=\fQ^\alpha{}_b
\bigl(v\Pi_{\rm b}+u\Pi_{\rm f}\bigr)~,\nonumber\\
\fS^a{}_{\beta,1}=\fS^a{}_\beta
\bigl(v\Pi_{\rm b}+u\Pi_{\rm f}\bigr)~,&\quad
\fS^a{}_{\beta,-1}=\fS^a{}_\beta
\bigl(u\Pi_{\rm b}+v\Pi_{\rm f}\bigr)~,
\end{align}
with $\Pi_{\rm b}$ ($\Pi_{\rm f}$) being the projector to bosons
(fermions) and
\begin{align}
u=\frac{1}{2}(x^++x^-)~,\quad v=\frac{1}{2}(1/x^++1/x^-)~.
\end{align}
The antipodes are given by 
${\rm S}(\fQ^\alpha{}_{b,\pm 1})=-\fU^{-1}\fQ^\alpha{}_{b,\pm 1}$ and
${\rm S}(\fS^a{}_{\beta,\pm 1})=-\fU^{+1}\fS^a{}_{\beta,\pm 1}$.
The names we have chosen for these operators are inspired by their
classical limit, and by the fact that they fulfill the crossing symmetry
property \eqref{antipoderel} at the quantum level.
Further commutators such as
$\{\Delta\fQ^\alpha{}_{b,1},\Delta\fS^a{}_\beta\}$ or
$\{\Delta\fQ^\alpha{}_b,\Delta\fS^a{}_{\beta,1}\}$ actually generate
the coproducts $\Delta\widehat\fR^a{}_b$ and
$\Delta\widehat\fL^\alpha{}_\beta$ of \cite{BeisYang}, which satisfy the
crossing symmetry relation at the quantum level, and coincide with the
generators $\Delta\fR^a{}_{b,1}$ and $\Delta\fL^\alpha{}_{\beta,1}$
found in \cite{MorTor} in the classical limit.

Let us list several further directions related to the results we have
presented here.

It is always crucial to study the symmetries when investigating the
correspondence between two theories.
The primary question which comes to mind is of course to see whether and
how the new one we have found in this paper will appear in the string
worldsheet theory.

An interesting application might be related to the string interaction
vertices.
In discussing the relation \cite{mst} between matrix string theory and
light-cone string field theory on the flat space, the supersymmetry
algebra on both sides plays an important role.
When it comes to the AdS/CFT correspondence on the pp-wave background at
the interacting level \cite{pp}, the story becomes more complicated.
It was questioned \cite{susy} whether supersymmetry alone can determine
the whole string interaction vertex \cite{SVPS}.
We would like to see whether our new symmetry can give further
restrictions in its construction.

Another application is related to the Hubbard model.
It was noticed \cite{Beinlin,MM} that the AdS/CFT S-matrix is equivalent
to Shastry's S-matrix.
Understanding how this new hidden symmetry appears in the Hubbard model
could help understanding more deeply the observed equivalence.

Finally, another interesting direction is to explore whether this
symmetry could be embedded in the full $\alg{psu(2,2|4)}$ algebra before
breaking it to $(\alg{su(2|2)})^2$, when the ground state is identified.
We reserve to return on these issues in future investigations.

\section*{Acknowledgments}
We would like to warmly thank G.~Arutyunov, H.~Awata, N.~Beisert,
S.~Benvenuti, H.Y.~Chen, S.~Frolov, J.~Gomis, H.~Kanno, Y.~Kazama,
T.~Klose, S.~Lee, X.~Liu, S.~Matsuura, T.~McLoughlin, J.~Minahan,
F.~Ravanini, S.~Sch\"afer-Nameki, K.~Yoshida and C.~Young for
interesting discussions.
A.T.\ especially thanks F.~Spill for many discussions and
for sharing his Diploma thesis.
S.M.\ is grateful for warm hospitality to the organizers of the
``Strings 007'' conference at Benasque, where part of this work was
done.
This work is supported in part by funds provided by the U.S. 
Department of Energy (D.O.E.) under cooperative research agreement
DE-FG02-05ER41360.
The work of S.M. is supported partly by Nishina Memorial Foundation,
Inamori Foundation and Grant-in-Aid for Young Scientists (\#18740143)
from the Japan Ministry of Education, Culture, Sports, Science and
Technology.
A.T. thanks Istituto Nazionale di Fisica Nucleare (I.N.F.N.) for
supporting him through a ``Bruno Rossi'' postdoctoral fellowship.

\end{document}